\documentclass[conference]{IEEEtran}
\IEEEoverridecommandlockouts
\usepackage{cite}
\usepackage{comment}
\usepackage{amsmath,amssymb,amsfonts}
\usepackage{algorithmic}
\usepackage{graphicx}
\usepackage{textcomp}
\usepackage{xcolor}
\usepackage{array}
\usepackage{multirow}
\usepackage{graphicx}
\def\BibTeX{{\rm B\kern-.05em{\sc i\kern-.025em b}\kern-.08em
    T\kern-.1667em\lower.7ex\hbox{E}\kern-.125emX}}
\begin{document}

\title{Deep Learning Models in Software Requirements Engineering}

\author{\IEEEauthorblockN{Maria Naumcheva}
\IEEEauthorblockA{\textit{Innopolis University}\\
Innopolis, Russia \\
m.naumcheva@innopolis.university}}

\maketitle

\begin{abstract}
Requirements elicitation is an important phase of any software project: the errors in requirements are more expensive to fix than the errors introduced at later stages of  software  life  cycle.  Nevertheless  many   projects do not devote sufficient  time  to  requirements. Automated requirements generation can improve the quality of software projects.
In this article we have accomplished the first step of the research on this topic: we have applied the vanilla sentence autoencoder to the sentence generation task and evaluated its performance.

The generated sentences are not plausible English and contain only a few meaningful words. We believe that applying the model to a larger dataset may produce significantly better results. Further research is needed to improve the quality of generated data.
\end{abstract}


\section{Introduction}

Machine learning (ML) is the field of study that gives computers the ability to learn without being explicitly programmed \cite{Samuel}. ML models allow to approach complex problems and large amounts of data. Instead of being explicitly programmed with  complex algorithms, they are trained on a training set to learn the patterns from the data. Machine learning has been applied to such tasks as anomaly detection \cite{rivera}, \cite{yakovlev}, car accident detection \cite {bortnikov}, \cite{batanina}, scene detection \cite{protasov}, image classification \cite{ahmad7}, hyperspectral image analysis and classification \cite{ahmad8}, \cite{ahmad9}, human activity recognition \cite{gavrilin2019across}, \cite{sozykin2018multi}, \cite{khan2010accelerometer}, \cite{khan2010triaxial}, object recognition \cite{khan2020post}, medical image analysis \cite{gusarev2017deep}, \cite{dobrenkii2017large}, machine translation \cite{khusainova2019sart}, \cite{khusainova2021hierarchical}, \cite{valeev2019application}.

Deep Learning models are ML models composed of multiple levels of non-linear operations, such as in neural nets with many hidden layers \cite{bengio2009learning}. Deep learning can be supervised, semi-supervised, or unsupervised. In supervised learning the data in the training set has labels, such as class name or target numeric value. Common supervised learning tasks are classification and regression (predicting target numerical value). In semi-supervised learning only part of the data is labeled. In unsupervised learning the data is unlabeled. Unsupervised learning is computationally more complex and solves harder problems, than the supervised learning models. 

Deep generative models are a rapidly evolving area of ML that enables modeling of such complex and high-dimensional data as text, images, and speech. The important examples of deep generative models are Generative Adversarial Networks (GAN) and Variational Autoencoders (VAE). GANs and VAEs have been applied to such tasks as image generation \cite{brock2018large}, image-to-image translation \cite{isola2017image}, sound synthesis \cite{engel2019gansynth}, data augmentation \cite{antoniou2017data}, music generation and natural language processing \cite{gui2020review}.

Requirements elicitation is an important phase of any software project. The errors in requirements are up to 200 times more expensive to fix than the errors introduced at later stages of software life cycle \cite{Boehm2001}. Nevertheless with the growing popularity of Agile software development, many projects quickly proceed to the implementation stage, without devoting sufficient time to requirements. Undocumented requirements pose threats to software project success: assumptions will have to be made during the development process, and they will not be agreed with the customer. If non-functional requirements are not documented, the developed software may have poor quality.

The automated requirements generation can solve the issue of poorly documented requirements as it may significantly reduce the time required for requirements elicitation. Automated requirements generation may involve extracting requirements from project documentation, conditional generation of non-functional requirements or machine translation of requirements in a programming language to a natural language.

Software requirements can be expressed in a natural language, in semi-formal notation (such as UML), in a graph- or automata-based notation, in mathematical notation, or in a programming language \cite{bruel2019}. According to the industrial survey \cite{fricker2015requirements}, 89\% of software projects specify requirements in natural language, so without losing generality we can say that requirements are texts. 

Out of deep generative models, unlike GANs, Variational Autoencoders can work with discrete data, such as texts, directly so they are a natural fit to text processing. Moreover they seem more suitable for the tasks of requirement generation as they are able to learn the semantic features and important attributes of the input text.

In our work we aim to do the first step in a journey to automated requirements generation. We apply the current state generative model to requirements generation task and outline the results. Although the application of VAE to natural language generation tasks is of interest for the research community, to the best of our knowledge the application of VAE in software requirements engineering domain has not yet been explored.

\section{Related Work}
\label{sec:related}
\subsection{NLP in Requirements engineering}

The applications of natural language processing (NLP) techniques to requirements engineering have been studied for decades. The systematic mapping study \cite{zhao2020natural} identifies such NLP tasks in requirements engineering, as: detecting linguistic issues, identifying domain abstractions and concepts, requirements classification, constructing conceptual models, establishing traceability links, requirements retrieval from existing repositories.
However, none of these tasks has been efficiently solved so far. Although the systematic mapping study identified 130 tools for requirements natural language processing, only 15 of them are available online and most of them are not supported anymore. 
One of the problems with applying the NLP techniques to requirements engineering is that there are no large or even medium-size datasets available. Creating datasets is costly, especially for labeled data. The deep generative models do not require annotated datasets so they may solve the problems intractable otherwise due to the lack of sufficient amount of labeled data.

\subsection{VAE for text generation}
In the sentence generating VAE, introduced by Bowman et al. \cite{bowman2015} the decoder performs sequential word generation, similar to the standard recurrent neural network language model (RNNLM), however the latent space also captures the global characteristics of the sentences, such as its topic or semantic properties. Yang et al. \cite{yang2017} empirically showed that using dilated CNN decoder, instead of LSTM decoder suggested by \cite{bowman2015}, can improve VAE performance for text modeling and classification and prevent posterior collapse.

Semi-supervised Sequential Variational Autoencoder (SSVAE), suggested by Xu et al. in \cite{xu2017}, targets the text classification task. The authors suggest feeding the input data label to the decoder at each time step. As a result, the classification accuracy significantly improves. Moreover, the trained model was able to generate plausible data: for the same latent variable z and different sentimental labels the model generated syntactically similar sentences with the opposite sentimental connotation.

Ml-VAE, introduced by Shen et al. \cite{shen2019} aims to cope with the task of long text generation. The multi-level LSTM structure of VAE is introduced in order to capture the text features at different levels of abstraction (sentence-level, word-level).

The task of conditional machine translation is explored in \cite{Zhang2016}, \cite{pagnoni2018}, and \cite{su2018}. Conditional VAE for machine translation models explicitly the input semantics and uses it for generating the translation. Ye et al \cite{ye2020variational} proposed a Variational Template Machine, a modification of vanilla VAE, for data-to-text generation. VTM has two disentangled latent variables: \textbf{$\mathsf{z}$}, modeling sentence template and \textbf{$\mathsf{c}$}, modeling content. 
Hu et al. address the problem of controlled text generation by combining VAE with attribute discriminators. In vanilla VAE the data is generated conditioned on the latent code $\mathsf{z}$ that has entangled dimentions. The proposed model combines $\mathsf{z}$ with the structured code $\mathsf{c}$, targeting sentence attributes. The resulting disentangled representation enables generation of sentences with given attributes \cite{hu2017}.

\section{Method}
\subsection{Model}
Variational Autoencoder, introduced in \cite{kingma2013}, is an unsupervised machine learning generative model, that consists of two connected networks, an encoder and a decoder. The encoder encodes the input as a distribution over the latent space, while decoder generates new data by reconstruction of data points generated from this distribution. The objective function has a form \cite{kingma2013}:

\begin{multline*}
\mathcal{L}(\theta, \phi; \mathsf{x^{(i)}}) = -D_{KL}(q_{\phi}(\mathsf{z}| | \mathsf{x}^{(i)}) || p_{\theta}(\mathsf{z}))\\ 
+ \dfrac{1}{L} \sum_{i=1}^L{(log{p_{\theta}(\mathsf{x}^{(i)} | \mathsf{z}^{(i,l)})})}
\end{multline*}

Our model is a sentence generating VAE, based on sequence to sequence RNN architecture \cite{bowman2015} \cite{blogTextVAE}. Recurrent Neural Networks are used to handle sequential data and can be utilized for inputs or outputs of variable lengths.  The model consists of bidirectional recurrent LSTM encoder, that maps the incoming sentence to latent random variables, normal RNN variational inference module and recurrent LSTM decoder that receives the latent representation as input at every step. 
The pre-trained GloVe word embeddings \cite{glove} are used for words vector representation.
The model architectural diagram is presented in Fig. 1.
\begin{figure}
\includegraphics[width=0.45\textwidth]{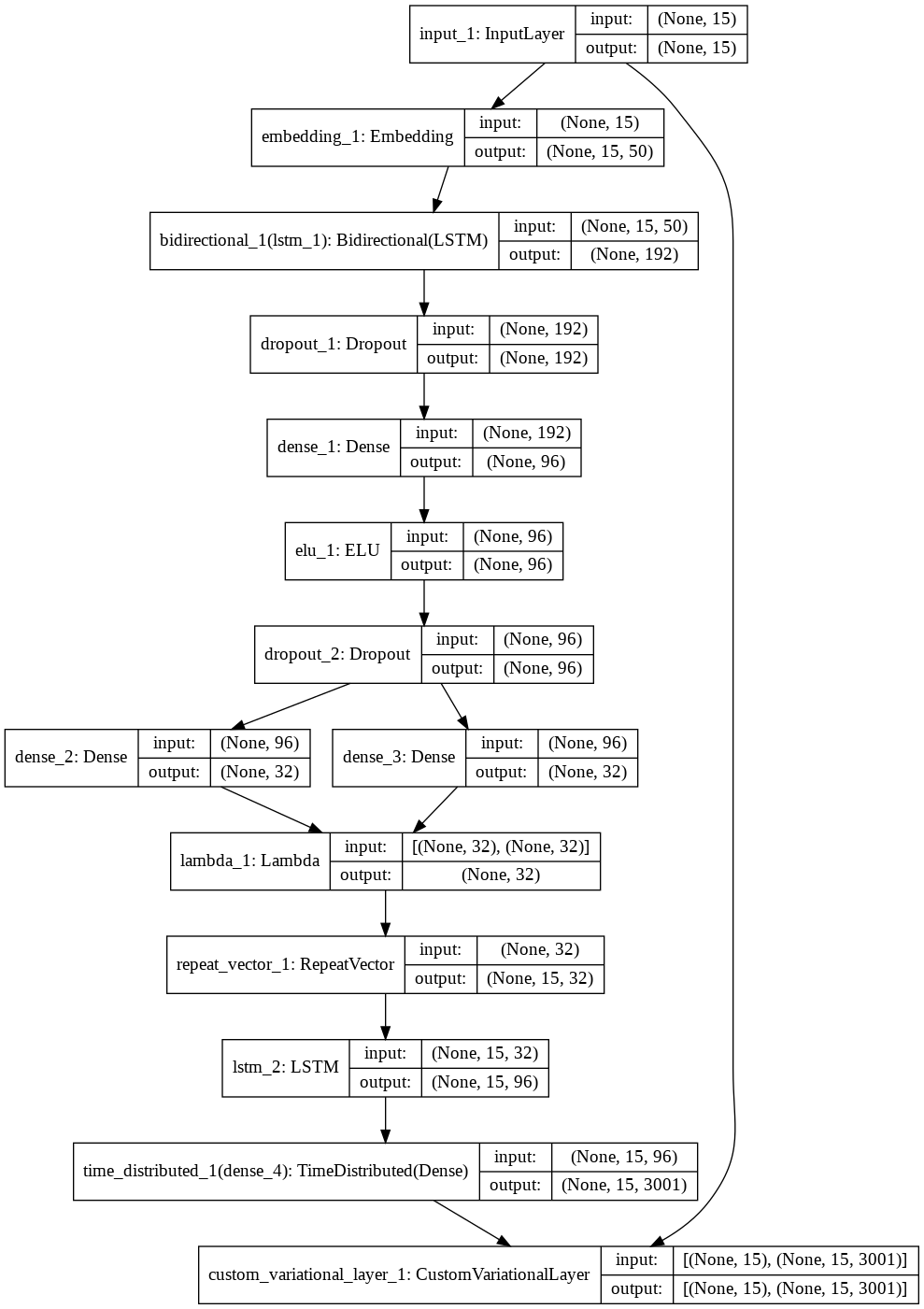}
\caption{Model architectural diagram}
\end{figure}
\subsection{Dataset}
Requirements texts have a specific structure. They are not the same as the sentences in fiction books or news article. One of the industry standards is so-called \textit{shall statements}, e.g. \textit{"The system shall notify the user when no matching product is found on the search."}. Some other requirements words, such as "should", "will", "must" may be used instead of "shall". Another industrial practise is to specify requirements as use cases. In the experiment section we do not consider use cases as requirements for the following reasons. First, requirements formulated as shall statements have common structure and each requirement (sentence) describes one atomic unit of functionality (or nonfunctional constraints), while some parts of use cases can not be treated as atomic requirements (such as for example stakeholders). Second, we believe that use cases are not a proper way to specify a system, they may serve as test cases for the requirements, but not as the requirements themselves (the detailed discussion on the topic is presented in \cite{meyerUML}).

For VAE implementation we used the requirements dataset, created for the ReqExp project, lead by Sadovykh A., Ivanov V., and Naumchev A., trimmed from the projects with poorly specified requirements and enriched with the data that we collected on our own from the publicly available software requirements specifications. The dataset has been checked for correctness, duplicated entries were removed, single requirements consisting of several sentences were split to several requirements, trimmed of excessive text or removed. The resulting dataset has about 3000 entries. Each entry is one sentence. The average length of one requirement is about 21 words. The example of dataset entries is listed in Table 1.
    
The data was preprocessed with Keras Tokenizer class. The Tokenizer  class turns each sentence into a sequence of integers (word indices in a dictionary). All punctuation is removed. The num\_words argument reflects the maximum number of words: only (num\_words - 1) most frequent words are kept.  

\begin{table}[h]
   \centering
  \begin{center}
 \begin{tabular}{|m{8.5cm}|}

 \hline
  \textbf{Requirements}\\ 
 \hline
The system should be able to create test environment of weborder system. \\
 \hline
The system hardware shall be fixed and patched via an internet connection. \\
 \hline
Yoggie shall coordinate on future enhancement and features with our organization. \\
 \hline
\end{tabular}
\end{center}

     \caption{Dataset excerpt}
    \label{fig:RQ}
\end{table}

\section{Evaluation and Future Work}
After training for 100 epochs, which took about 1.5 hours, we checked the ability of trained model to generate requirements. To do so, we interpolated in the latent space between two requirements. The generated sentences are presented in Table 2

\begin{table}[h]
    \centering
    \begin{center}
 \begin{tabular}{|m{1.5cm}|m{6.0cm}|} 
 \hline
  real sentence & \textbf{Customers shall be able to confirm the order after checkout}\\ 
 \hline
 generated sentence & the system shall shall shall to to to to the the the \\
 \hline
 real sentence & \textbf{Users shall be able to turn the sound on or off at any stage during the game}\\
 \hline
 generated sentence & the shall shall to to the the the the the the the the the the \\
 \hline
 real sentence & \textbf{The system must allow only admin-users to set up user profiles and allocate users to groups} \\
 \hline
 generated sentence & the shall shall to the the the the the the the the the the the \\
 \hline
 real sentence & \textbf{The system should support multilingual interface} \\
 \hline
 generated sentence & the shall shall shall the the \\
 \hline
 real sentence & \textbf{the system should be developed on open standards} \\
  \hline
 generated sentence & the system shall shall shall to to to the the \\
 \hline
\end{tabular}
\end{center}

    \caption{Requirements generated by the model}
    \label{fig:RQ}
\end{table}
As we see, the only meaningful words that appeared in the generated requirements are "system" and "shall". The rest are stop words "the" and "to". At the same time, "system" and "shall" are indeed the most important words in software requirements. As the dataset was relatively small (about 3000 entries), we believe that better results could be achieved on a substantially larger dataset. Nevertheless we have to admit that the current results of the model are not satisfactory.

We see the following future work directions. First and most important is to collect a substantially larger software requirements dataset. Second, experiment with the existing model on a larger dataset. Third, experiment with another VAE models, such as ml-VAE, conditional VAE, Variational Template Machine, and VAE for controlled text generation.

\bibliographystyle{unsrt}  
\bibliography{references}

\end{document}